\documentclass[conference]{IEEEtran}
\IEEEoverridecommandlockouts

\usepackage{listings}
\usepackage{tcolorbox}
\usepackage[normalem]{ulem}
\usepackage{colortbl}
\usepackage{xcolor}
\usepackage[table]{xcolor}
\usepackage{booktabs}
\usepackage{balance}

\definecolor{lightpurple}{RGB}{235,230,250}
\definecolor{lightgreen}{RGB}{235,250,240}

\usepackage{caption}
\AtBeginDocument{%
  \providecommand\BibTeX{{%
      Bib\TeX}}}

\lstdefinelanguage{Python}{
    morekeywords={class, def, return, try, except, raise, from},
    keywordstyle=\color{blue},
    stringstyle=\color{green},
    commentstyle=\color{gray},
    morecomment=[l]{\#},
}

\lstdefinelanguage{diff}{
    morecomment=[f][\color{green}]{+},
    morecomment=[f][\color{red}]{-},
    morecomment=[f][\color{blue}]{@@},
}

\lstdefinelanguage{errorlog}{
    morecomment=[f][\color{red}]{E},
    morecomment=[f][\color{magenta}]{?},
}

\usepackage{algorithm}
\usepackage{algorithmic}

\usepackage{listings}
\usepackage{xspace}

\usepackage[T1]{fontenc}

\usepackage[utf8]{inputenc}

\usepackage{microtype}

\usepackage{graphicx}

\usepackage{subcaption}
\usepackage{fontawesome5}

\usepackage{soul}
\usepackage{framed}
\usepackage{multirow}
\usepackage{siunitx}
\usepackage{float}

\newcommand*\colourcheck[1]{%
	\expandafter\newcommand\csname #1check\endcsname{\textcolor{#1}{\ding{52}}}%
}
\colourcheck{blue}
\colourcheck{green}
\colourcheck{red}

\definecolor{custom-blue}{rgb}{0,0,0}

\newcommand{\tool}{SpecTune\xspace}

\def\BibTeX{{\rm B\kern-.05em{\sc i\kern-.025em b}\kern-.08em
    T\kern-.1667em\lower.7ex\hbox{E}\kern-.125emX}}

\newboolean{showcomments}
\setboolean{showcomments}{true}
\ifthenelse{\boolean{showcomments}}
 { \newcommand{\mynote}[2]{
      \fbox{\bfseries\sffamily\scriptsize#1}
        {\small$\blacktriangleright$\textsf{\emph{#2}}$\blacktriangleleft$}}}
        { \newcommand{\mynote}[2]{}}

\usepackage{tikz}

\newcolumntype{L}[1]{>{\raggedright\arraybackslash}p{#1}}

\newcommand{\code}[1]{{\footnotesize\texttt{#1}}}
\usepackage{amsthm}
\definecolor{dkgreen}{rgb}{0,0.6,0}
\definecolor{gray}{rgb}{0.5,0.5,0.5}
\definecolor{lightgray}{rgb}{211, 211, 211}
\definecolor{mauve}{rgb}{0.58,0,0.82}
\definecolor{lightblue}{RGB}{245, 245, 220}
\definecolor{grey}{RGB}{169,169,169}
\definecolor{darkgreen}{RGB}{54, 156, 90}
\definecolor{lightblue}{RGB}{235, 247, 255}
\definecolor{niceblue}{RGB}{52, 164, 235}
\definecolor{nicered}{RGB}{219, 42, 48}
\definecolor{niceorange}{RGB}{224, 114, 54}
\definecolor{lightred}{RGB}{255, 227, 228}
\definecolor{lightorange}{RGB}{252, 235, 227}
\definecolor{orange1}{RGB}{214, 178, 131}
\definecolor{brown}{RGB}{163, 117, 57}
\definecolor{lightbrown}{RGB}{224, 163, 83}
\definecolor{commentcolor}{RGB}{101, 163, 178}
\definecolor{lightpurple}{RGB}{203, 195, 227}

\definecolor{custom-blue}{rgb}{0,0,0}

\definecolor{c1}{HTML}{f4cccc}
\definecolor{c2}{HTML}{f5cdcd}
\definecolor{c3}{HTML}{fffcfc}
\definecolor{c4}{HTML}{ffffff}
\definecolor{c5}{HTML}{ffffff}
\definecolor{c6}{HTML}{fffdfd}
\definecolor{c7}{HTML}{f5cfcf}
\definecolor{c8}{HTML}{fffbfb}
\definecolor{c9}{HTML}{ffffff}
\definecolor{c10}{HTML}{fffdfd}
\definecolor{c11}{HTML}{fefafa}
\definecolor{c12}{HTML}{fef7f7}
\definecolor{c13}{HTML}{ffffff}
\definecolor{c14}{HTML}{fffefe}
\definecolor{c15}{HTML}{ffffff}
\definecolor{c16}{HTML}{fefafa}
\definecolor{c17}{HTML}{fdf3f3}
\definecolor{c18}{HTML}{fffefe}
\definecolor{c19}{HTML}{fdf5f5}
\definecolor{c20}{HTML}{ffffff}

\definecolor{NonConsistent}{HTML}{DC2626} 
\definecolor{Trivial}{HTML}{595F6B} 
\definecolor{Acceptable}{HTML}{156127}

\makeatletter
\newcommand{\linebreakand}{%
  \end{@IEEEauthorhalign}
  \hfill\mbox{}\par
  \mbox{}\hfill\begin{@IEEEauthorhalign}
}
\makeatother

\begin{document}

\title{Enhancing Program Repair with Specification Guidance and Intermediate Behavioral Signals}


\author{
\IEEEauthorblockN{Minh Le-Anh\textsuperscript{*}}
\IEEEauthorblockA{\textit{FPT Software AI Center} \\
\textit{Hanoi Univ. of Science and Tech.} \\
Hanoi, Vietnam \\
minhla4@fpt.com}
\and
\IEEEauthorblockN{Cuong Chi Le\textsuperscript{*}}
\IEEEauthorblockA{\textit{University of Texas at Dallas} \\
Texas, USA \\
cuong.le@utdallas.edu}
\and
\IEEEauthorblockN{Tien N. Nguyen}
\IEEEauthorblockA{\textit{University of Texas at Dallas} \\
Texas, USA \\
tien.n.nguyen@utdallas.edu}
\thanks{\textsuperscript{*}Equal Contribution}
}

\maketitle

\begin{abstract}
Automated Program Repair (APR) has recently benefited from large language models (LLMs). However, most LLM-based APR approaches still rely primarily on coarse-grained, end-to-end signals from test-suite outcomes or specification checks on the candidate fix to guide repair, providing limited insight into where a program’s internal logic deviates from its intended behavior. In contrast, human debugging often relies on intermediate reasoning about program states through localized correctness conditions or assertions. Inspired by this observation, we propose {\tool}, a specification-guided debugging framework that incorporates intermediate behavioral reasoning into APR. {\tool} decomposes the repair task~into suspicious regions connected by execution checkpoints and derives localized postconditions representing expected program behaviors at those points. By executing the buggy program and evaluating these postconditions, {\tool} produces micro-level~debugging signals that indicate mismatches between observed and intended behaviors, enabling more precise fault localization and targeted patch generation. To address the potential unreliability of LLM-generated postconditions, we introduce two complementary signals: a specification validation signal $\alpha$, which estimates the consistency of generated postconditions using partial passing test cases, and a discriminative signal 
$\beta$, which detects violations of validated postconditions during execution. With these signals, {\tool} safely leverages automatically generated specifications for APR. Experimental results show that {\tool} improves fault localization and APR effectiveness than the baselines.

\end{abstract}


\section{Introduction}
\label{sec:intro}

Automated Program Repair (APR) has evolved through several methodological paradigms. Early work largely followed the \emph{generate-and-validate} paradigm, where candidate patches are enumerated and validated against a test suite serving as the oracle. Search-based approaches~\cite{weimer2009genprog,le2011genprog} explored syntactic edits via evolutionary mutation, while template-driven and learning-guided systems~\cite{long2015spr,long2016prophet} constrained or ranked candidate patches using repair templates and properties of human-written fixes. Subsequently, semantic and pattern-based approaches~\cite{nguyen2013semfix,xuan2017nopol,mechtaev2016angelix,kim2013par,koyuncu2018fixminer} improved repair by leveraging symbolic execution, constraint solving, execution traces, and recurring fix patterns. Later, deep learning approaches~\cite{chen2019sequencer,lutellier2020coconut,icse20-dlfix,icse22-dear} reframed repair as translation from buggy to fixed code, learning repair patterns directly from bug-fixing datasets.

The advances of large language models (LLMs) has transformed APR. Modern LLM-based APR approaches
can be broadly grouped into the following categories. {\bf First,} {\em one-shot prompting} with validation workflows can help LLMs outperform traditional APR systems~\cite{xia2023plm_apr,xia2024automated,fan2023apr_llm_outputs}. This line of
approaches relies mainly on the {\em APR capability of LLMs}. {\bf Second,} recent approaches further enhance repair through context {\em retrieval augmentation} (e.g., RAP-Gen~\cite{wang2023rapgen}). {\bf Third,} {\em iterative repair strategies} formulate APR as an iterative process that selectively refines promising candidate trajectories
with different types of feedback (e.g., ChatRepair~\cite{xia2024automated}, REx~\cite{tang2024code}). {\bf Fourth}, advanced approaches enhance repair through {\em multi-agent frameworks} (e.g., AutoCodeRover~\cite{zhang2024autocoderover}, RepairAgent~\cite{bouzenia2025repairagent}, AgentCoder~\cite{huang2024agentcodermultiagentbasedcodegeneration}) that iteratively localize faults and refine patches using external tools (e.g., compilation and testing). 
 
Despite these advances, existing APR approaches still rely primarily on {\bf \emph{macro-level debugging signals}}: they~evaluate candidate repairs based on final outcomes, such~as test-suite pass/fail results~\cite{tang2024code}. Other approaches explore~specification signals in different forms. For example, SpecRover~\cite{ruan2025specrover}~uses natural-language specifications to guide and vet patches in AutoCodeRover~\cite{zhang2024autocoderover} agentic APR framework, while~nl2postcond~\cite{nl2postcond} studies the generation of formal postconditions from natural-language intent. However, these signals are typically method-level or candidate-level, rather than checkpoint-level behavioral signals inside the buggy execution.


While effective, such signals are coarse-grained. A failing test or a failing check with a specification indicates that the candidate's output/logic is incorrect, but it often provides limited {\bf \em fine-grained} information about {\em where the program's internal state first deviates from the intended behavior}. Consequently, the APR process frequently degenerates into a large search over candidate edits guided only by coarse-grained signals.

This reliance on macro signals does not reflect how developers {\em debug programs}. Human debugging often proceeds by decomposing the computation into smaller parts and reasoning about intermediate program states at {\bf \em checkpoints}. Developers form {\bf \em hypotheses} about what the program \emph{should satisfy} at certain checkpoints and validate these hypotheses through intermediate checks. When the observed behavior violates such expectations, the developer can narrow the fault scope and modify the responsible code region. In essence, human debugging is often guided by {\bf \emph{intermediate behavioral specifications}} rather than only by end-to-end pass/fail outcome.
This suggests a new direction for APR: leveraging intermediate specifications as structured debugging signals.

We propose {\bf {\tool}, a \emph{specification-guided debugging framework} for APR} that incorporates intermediate behavioral reasoning into the APR process. In a {\bf \em single-pass setting}, it aims to make an LLM more effective than the same LLM used alone. In an {\bf \em iterative repair setting}, the goal is to improve an iterative refinement framework (without {\tool}), by providing intermediate behavioral guidance. We also aim for it to improve a {\bf \em file-level, agentic} APR framework. These comparisons isolate whether \tool adds value to the same repair model or iterative framework. We do not target {\bf \em repo-level}, multi-agent APR systems, since they introduce additional confounding factors, including context retrieval, tool use, planning, test generation, build setup, and patch ranking, making it difficult to attribute improvements to \tool.

Given a buggy program and its problem description, the target LLM is prompted to decompose the repair task into suspicious regions separated by intermediate checkpoints.~For each checkpoint, it is prompted to derive a localized postcondition that captures the expected program behavior at that point. The buggy program is then executed via the given test cases and these postconditions are evaluated against the observed program states. When a specification is violated, the resulting mismatch indicates that the program logic near that checkpoint may be faulty. 
These mismatch signals provide structured guidance for fault localization (FL) and targeted fix generation.


%
The first key challenge is that these LLM-generated postconditions are not always reliable. 
In our work, we intentionally keep the generation strategy simple and focus on controlling specification quality after generation. Second, poorly placed checkpoints are usually not useful for repair as they produce postconditions that are either inconsistent with passing executions or not discriminative on failing executions.  Thus, our tool validates and filters generated specifications before using them via two complementary signals. The first is a {\bf \emph{specification validation signal}} $\alpha$, which addresses the potential unreliability of LLM-generated postconditions. The $\alpha$ signal is computed using partial passing test cases obtained from executing the original program and estimates whether a generated postcondition is consistent with the intended behavior supported by available evidence. The second is a {\bf \emph{micro-level discriminative signal}} $\beta$, 
which estimates how often a postcondition remains satisfied on failing executions; lower $\beta$ indicates a stronger ability to expose behavioral mismatches. This paper makes the following contributions:


\begin{figure}[t]
    \centering
    \includegraphics[width=3.6in]{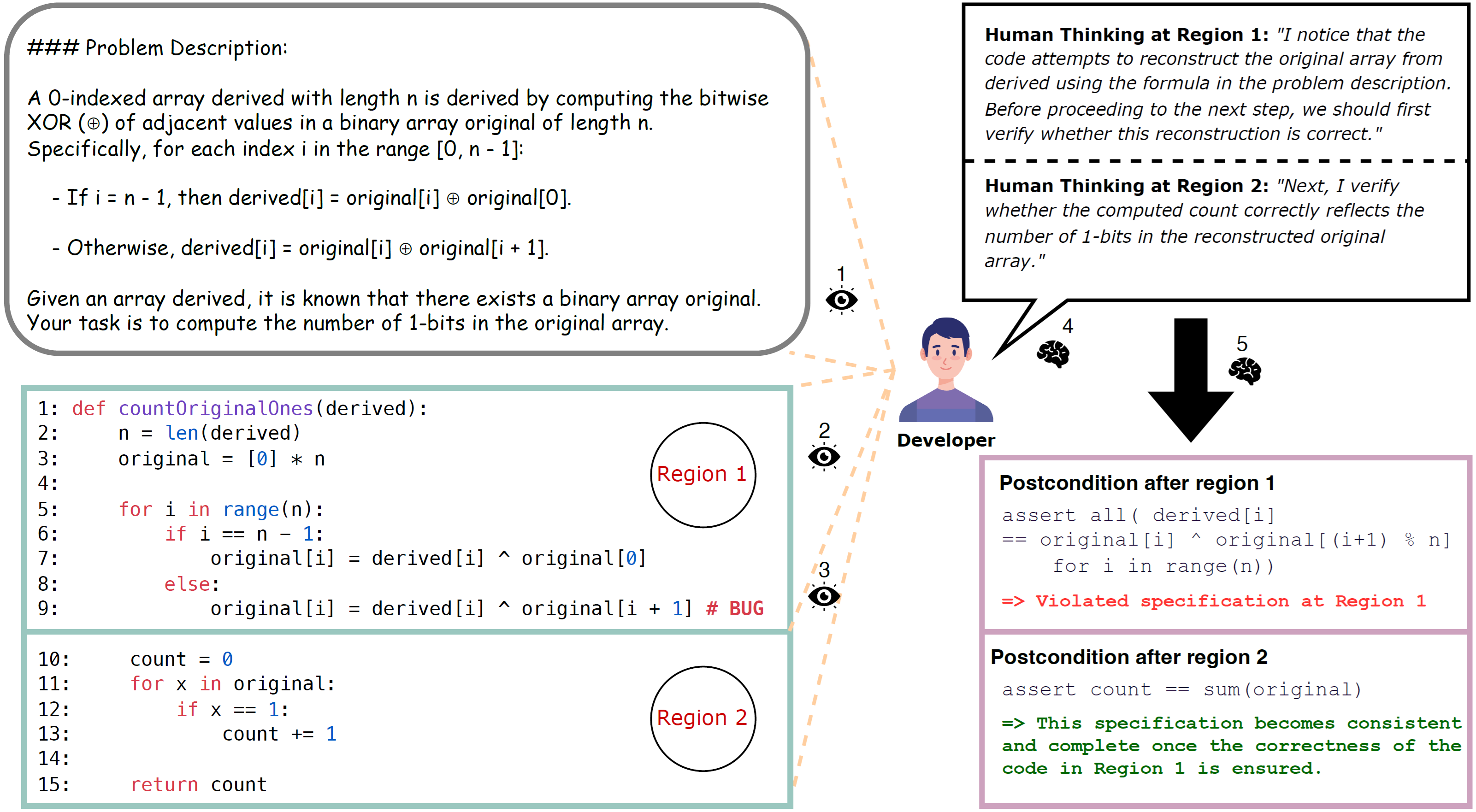}
    \vspace{-9pt}
    \captionsetup{font=footnotesize}`
    \caption{
    A developer derives debugging signals from program semantics.
\textbf{Step 1:} Read the problem description to understand the intended behavior.
\textbf{Step 2:} Inspect relevant code regions and reason about expected variable relationships and formulate hypotheses as postconditions.
\textbf{Step 4:} Evaluate the postconditions in execution to reveal mismatches between actual and intended behavior.
        }
    \label{fig:motiv}
\end{figure}



(1) {\tool}, a specification-guided APR framework that decomposes APR into checkpoint-based reasoning steps and leverages intermediate postconditions to guide repair. 


(2) We introduce checkpoint-level postcondition violations as micro-level debugging signals for guiding LLM-based APR.

(3) We design a dual-signal filtering algorithm that uses $\alpha$ to remove unreliable postconditions and $\beta$ to remove weak or non-discriminative postconditions.

(4) Empirical evaluation, showing that {\tool} can enhance APR performance 
over the state-of-the-art approaches.
\section{Motivation}
\label{sec:motivation}


Fig.~\ref{fig:motiv} illustrates an example.
Firstly, the intended program behavior is given in the problem description (\faEye\ 1). Given the buggy code in the bottom left corner of Figure~\ref{fig:motiv}, a developer will 
identify one or more critical code regions and reason about the expected relations between intermediate variables during execution (\faEye\ 2, \faEye\ 3). For example, in Fig.~\ref{fig:motiv}, once the developer understands that the task is to reconstruct the array \code{original} from the XOR array \code{derived} and count the number of `1's, they inspect the current code. In particular, Region~1 corresponds to the reconstruction step that attempts to recover the \code{original} array  from the \code{derived} array, while Region~2 performs the counting step that computes the number of `1's in the reconstructed array. Instead of attempting repair the entire buggy program at once, (s)he typically validates each code region. For example, (s)he could form a hypothesis at the end of Region 1 (\faBrain 4). From the perspective of a model, a formulation of such hypothesis could be expressed as an executable postcondition that captures the expected program state after Region 1 (\faBrain 5). In our example, the postcondition is whether the reconstructed array \code{original} satisfies the XOR relationship with the \code{derived} array:
\[\scalebox{0.88}{$
\forall i \in [0, n-1],\;
\code{derived}[i] =
\code{original}[i] \oplus
\code{original}[(i+1) \bmod n]
$}
\]
However, when implementing the reconstruction logic, the \code{original} array is initialized with zeros. During the iterative reconstruction process, the elements of the \code{original} array are expected to be computed sequentially, where each newly reconstructed value depends on previously computed elements. In particular, when reconstructing the array, the value of \code{original[i+1]} should be derived from the already computed element \code{original[i]}. However, the buggy code in Figure~\ref{fig:motiv} instead uses \code{original[i+1]} when updating \code{original[i]}, which introduces a forward dependency on a value that has not yet been correctly derived. This leads to the bug. 


The actual/observed behavior of the buggy program is given through the passing/failing test cases, while the intended behavior is formulated as a postcondition at each checkpoint. The mismatch between the two behaviors at a checkpoint indicates that the logic leading to that checkpoint might be faulty. Overall, decomposing a buggy program into suspicious functional regions and validating each region with dedicated postconditions can help a human or an LLM significantly narrow the search space for bug localization and repair. By focusing on localized behavioral constraints, developers can reason about the correctness of individual regions.
Importantly, from the mismatch, the developer could fix the buggy code. 

\section{Key Ideas}
\label{sec:key_ideas}



\vspace{2pt}
\noindent {\bf [Key Idea 1] \em Specification-guided debugging.}
In practice, developers typically decompose the overall problem and buggy code into multiple smaller regions, progressively narrowing down the search space to localize and fix bugs. Each region is then validated using local unit checks, such as assertions or intermediate correctness conditions. When a violation is observed, developers fix the corresponding code region to restore the expected behavior.

$\hookrightarrow$ Inspired by this debugging strategy, we propose \tool, a specification-guided debugging framework that decomposes the repair problem into a sequence of suspicious regions~separated by intermediate {\bf checkpoints}. For a checkpoint, it derives a {\bf postcondition} that aligns with the intended program behavior mentioned in the problem description. By executing the program and evaluating these specifications, it aims to identify {\bf mismatches} between the observed program behavior and the intended specifications. {\em These mismatch signals provide structured guidance for localizing faults and generating fixes}. 

\begin{figure*}
    \centering
    \includegraphics[width=0.85\textwidth]{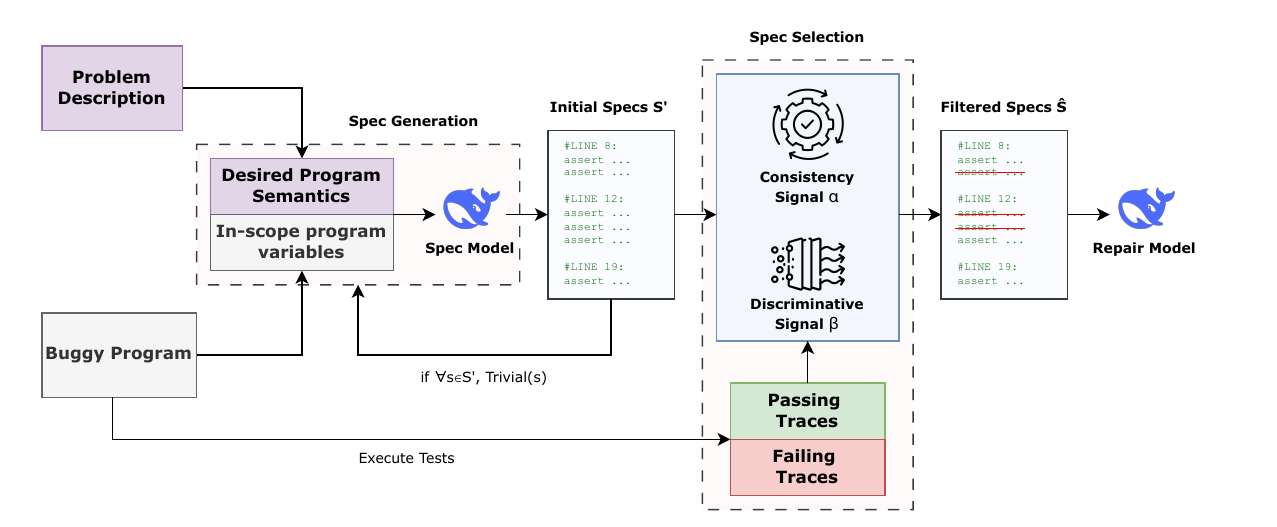}
    \vspace{-6pt}
    \captionsetup{font=footnotesize}
    \caption{{\tool} framework starts with a {\bf \em buggy program, a test suite, and its problem description}, from which candidate specifications are generated based on program semantics and in-scope variables. These specifications are then validated using passing and failing test executions to compute the consistency ($\alpha$) and discriminative ($\beta$) signals, respectively. After filtering, the refined specification set $\hat{S}$ is used to guide the repair model in generating corrected programs.}
    \label{fig:overview}
\end{figure*}

\vspace{3pt}
\noindent {\bf [Key Idea 2]. \em Micro Debugging Signals.}
Existing APR approaches primarily rely on coarse-grained debugging signals, i.e., {\em assess the repair candidates as a whole} via test cases~\cite{xia2024automated} or checking them with natural-language specifications~\cite{ruan2025specrover}. 
In contrast, the intrinsic logic of a program is typically validated through finer-grained checks, such as intermediate assertions or localized correctness conditions. These micro-level tests verify whether individual functions or intermediate states of the program satisfy specific logical properties. 

$\hookrightarrow$ To derive intermediate specifications, we leverage large language models (LLMs) to infer postconditions at selected execution checkpoints. Given the problem description and the surrounding program context, an {\em LLM generates the checkpoints and postconditions} that capture the expected properties and relationships among the relevant variables. These postconditions describe the intended program state after the execution to the corresponding code region. Prior work~\cite{nl2postcond} has demonstrated that LLMs can effectively synthesize 
postconditions and assertions from textual descriptions or source code, making them suitable for generating intermediate behavioral constraints.


\vspace{3pt}
\noindent \textbf{[Key Idea 3] \em Specification Validation.}
Directly relying on these specifications from LLMs may lead to incorrect localization due to their hallucinations/imprecision. To~address this, we introduce a {\bf correctness validation signal} $\alpha$, computed from the partial passing test cases gathered through running the~code. This acts as a validation to quantify the {\bf correctness} of~ge\-nerated postconditions based on their consistency with the program behavior during the executions of passing test cases.

For each generated postcondition $p$, we also design a {\bf discriminative signal $\beta$}, a type of micro-level signal that {\em assesses the quality of the generated postcondition $p$ on its ability to distinguish incorrect behavior (in failing test cases) from the intended behavior (via $p$)}. The postconditions with high discriminative signals $\beta$ are treated as potential indicators of faulty logic and used as contextual guidance for the LLM in APR.
In Fig.~\ref{fig:motiv}, the mismatch between the observed program state at the end of Region~1 under failing test executions and the inferred postcondition~1 (right-hand side) constitutes a micro-level debugging signal. That is, $\beta$ assesses the discriminative characteristic (also referred to as {\bf completeness}) of a generated postcondition.
It also helps remove the `weak' postcondition (e.g., \code{`true} == \code{true'}) that cannot detect a bug.
\section{{\tool} Approach}
\label{sec:approach}

\subsection{Problem Formulation}

We formulate the specification-driven APR problem as~follows. Given a buggy program $C$, a  description, and a test suite $T$ = $\{t_1, t_2, \ldots, t_n\}$, where each test case $t_i$ = $(x_i, y_i)$ consists of an input $x_i$ and its expected output $y_i$ produced by the canonical program $C^*$. By executing the buggy program $C$ on each test case in $T$, we partition the test suite into two subsets: the passing test cases $P = \{\, t_i \in T \mid C(x_i) = y_i \,\}$ and the failing ones $F = \{\, t_j \in T \mid C(x_j) \neq y_j \,\}$. Our objective is to generate a set of postconditions $S$ that are both consistent with and complete with respect to the intended program behavior $I$, which is specified in the description. Ideally, we aim to find an optimal condition set $S^*$ that maximizes completeness while satisfying a sufficiently high consistency requirement:
\[\scalebox{0.94}{$
S^* =
\arg\max_S \text{Completeness}(S,I)
\quad
\text{s.t.}
\quad
\text{Consistency}(S,I) \ge \theta
$}
\]
where $\text{Consistency}(S, I)$ measures how well the postconditions $S$ align with the intended behavior $I$, and $\text{Completeness}(S, I)$ measures $S$'s \text{discriminative} ability to expose deviated behaviors in incorrect mutated executions, rather than being~trivial.

Directly measuring \textbf{Consistency} and \textbf{Completeness} is infeasible in practice, since both depend on the correct program $C^*$ and its intermediate execution states, which are unavailable in APR. Thus, we introduce \textit{$\alpha$-$\beta$ algorithm} that assesses the quality of generated postconditions using only observable executions of the buggy program. Specifically, the $\alpha$ signal leverages the passing test cases $P$ to estimate whether a generated postcondition is consistent with the intended behavior, while the $\beta$ signal leverages failing executions to estimate whether the postcondition is sufficiently discriminative.

Since \tool operates over a finite set (S') of LLM-generated candidate postconditions, it does not solve a global optimization problem. Instead, it realizes the objective through threshold-based selection over (S'). The threshold on $\alpha$ serves as a reliability filter, removing postconditions that are not supported by passing executions. The threshold on $\beta$ acts as a discriminative filter, removing postconditions that are too weak, incomplete, or uninformative for debugging.



The refined specification set $\hat{S}$ is then used to guide the LLM. Let $M_{\phi}$ denote a LLM parameterized by $\phi$. The repair process can be formulated as generating a candidate fixed program $\hat{C}$ from the conditional distribution: $\hat{C} \sim M_{\phi}(\cdot \mid C, \hat{S})$, 
where $\hat{S}$ provides structured behavioral guidance that steers the model toward repairs consistent with the intended program semantics.


\subsection{{\tool} Workflow}



{\tool}'s workflow is illustrated in Fig.~\ref{fig:overview}.
{\tool} takes~as input a buggy program $C$, its functional description $D$, a test suite~$T$, together with the consistency and discriminative thresholds $\theta$ and~$\gamma$. The output is the corrected program.



\subsubsection{\bf Checkpoints and Postcondition Generation}
{\tool} first prompts an LLM to construct candidate postconditions at multiple LLM-generated checkpoints. We leverage the capability of Code LLMs to analyze and reason over program logic to identify a set of intermediate checkpoints $Q = \{q_1, q_2, \ldots, q_n\}$ within the buggy program. These checkpoints correspond to critical execution points where the internal program state can reveal potential logical inconsistencies. For each checkpoint $q_j$, a set of candidate postconditions $S'(q_j) = \{s_{j0}, s_{j1}, \ldots, s_{jk}\}$  is associated, where each specification is expected to hold immediately after execution reaches that checkpoint. 
To facilitate checkpoint and postcondition generation, we provide the LLM with contextual information, including a failing input-output test case of the buggy program. Conditioned on this information, we obtain a set of candidate postconditions for each checkpoint, forming the initial specifications $S' = \bigcup_{q_j \in Q} S'(q_j)$. {\tool} executes the code on both passing and failing tests and collects traces and program states at each checkpoint.

\subsubsection{\bf Specification Validation}
Although LLMs show promising capability in proposing candidate specifications~\cite{nl2postcond}, {\em it remains necessary to verify whether these generated postconditions are consistent with the intended behavior and sufficiently~informative for debugging}. 
Concretely, let $P = \{p_1, p_2, \ldots, p_m\}$ and $F = \{f_1, f_2, \ldots, f_l\}$ denote the sets of passing and failing test cases, respectively. Let $Q = \{q_1, q_2, \ldots, q_n\}$ denote the set of checkpoints within the buggy program proposed by LLMs. For each checkpoint $q_j$, we have an initial set of candidate LLM-generated postconditions $S'_j = \{s_{j1}, s_{j2}, \ldots, s_{jk}\}$. From these, we compute two signals $\alpha$ and $\beta$ to assess the quality of candidate conditions:

\vspace{1pt}
\textbf{Consistency Signal ($\alpha$).}
For each candidate specification $s_{jk}$ associated with checkpoint $q_j$, we approximate its consistency with the intended program behavior using the $\alpha$ signal computed over the passing test cases. {\em The consistency signal $\alpha$ of a postcondition $s_{jk}$ is defined as the probability that the specification holds when a passing test reaches the corresponding checkpoint}:
\begin{equation}
\alpha(s_{jk}) =
\frac{
|\{\, p_i \in P \mid p_i \text{ reaches } q_j \land x_{ij} \models s_{jk} \,\}|
}{
|\{\, p_i \in P \mid p_i \text{ reaches } q_j \,\}|
}
\end{equation}
where $x_{ij}$ denotes the program state observed when executing test case $p_i$ at checkpoint $q_j$.
Based on this signal, we filter out specifications with low $\alpha$ values, as they are likely inconsistent with the intended program behavior. Specifically, we apply a threshold $\theta \in [0,1]$ and retain only the specifications whose consistency signals satisfy $\alpha(s_{jk}) \geq \theta$.

\vspace{1pt}
\textbf{Discriminative Signal ($\beta$).}
Evaluating consistency alone is insufficient, as some LLM-generated postconditions may be trivially satisfied and fail to capture erroneous behaviors due to a poorly placed checkpoint. Thus, we introduce {\em the signal $\beta$ to assess whether a postcondition $s_{jk}$ can effectively identify incorrect executions using failing test cases}:
\begin{equation}
\beta(s_{jk}) =
\frac{
|\{\, f_i \in F \mid f_i \text{ reaches } q_j \land y_{ij} \models s_{jk} \,\}|
}{
|\{\, f_i \in F \mid f_i \text{ reaches } q_j \,\}|
}
\end{equation}
A higher $\beta$ indicates that the postcondition is frequently satisfied during the executions of failing test cases. This may arise either because the postcondition is incomplete and fails to capture the erroneous behavior, or because the failing execution does not reach the program state where the logical mismatch manifests. In contrast, postconditions with lower $\beta$ signals exhibit stronger discriminative capability, as they are more likely to be violated by failing executions and give more informative signals for exposing incorrect program behaviors. To remove incomplete or uninformative postconditions for APR, we filter candidates using a threshold $\gamma \in [0,1]$ and retain only those whose discriminative signals satisfy $\beta(s_{jk}) < \gamma$.

\subsubsection{\bf Specification Selection}
Overall, the postconditions satisfying both the consistency ($\alpha_{jk} \ge \theta$) and discriminative ($\beta_{jk} < \gamma$) criteria are added to the refined specification set $\hat{S}$, while the others are discarded,
ensuring that the resulting postconditions are both consistent with correct executions and capable of exposing erroneous behaviors. 
\begin{equation}
\hat{S} =
\{\, s_{jk} \in S' \mid \alpha(s_{jk}) \geq \theta \;\wedge\; \beta(s_{jk}) < \gamma \,\}
\end{equation}
Finally, the LLM repair model is prompted to generate a candidate fixed program $\hat{C}$ conditioned on the buggy program $C$ and the validated specification set $\hat{S}$.

\vspace{-4pt}
\section{Empirical Evaluation}
\label{sec:exp}

For evaluation, we seek to answer the following questions:

\textbf{RQ1. [Program Repair Performance]} What is {\tool}'s program repair performance across different settings?

\textbf{RQ2. [Fault Localization Capability]} 
How does \tool compare with existing baselines in fault localization?


{\bf RQ3. [Effectiveness of $\alpha$-$\beta$ Algorithm].} How effective is our algorithm
in keeping the high-quality postconditions?

\textbf{RQ4. [Ablation Study]} What is the individual contribution of each signal in SpecTune?

\textbf{RQ5. [Sensitivity Analysis]} 
How sensitive is the performance of \tool to the choices of the $\alpha$ and $\beta$ thresholds?

Our evaluation is designed to isolate the contribution of~{\tool} under its intended scope. Since {\tool} {\em provides~intermediate behavioral guidance} rather than a full repo-level repair workflow, {\em we attach it to the same repair frameworks used by the state-of-the-art baselines}. This allows us to {\bf \em compare \tool with other specification feedback signals/baselines} in one-turn LLM repair, iterative refinement repair, and file-level agentic repair frameworks/settings while keeping the framework/setting fixed. {\em We do not compare with full repo-level multi-agent APR systems as end-to-end baselines}, since their performance depends on additional components such as repository search, tool use, build setup, and patch ranking, making it difficult to attribute gains specifically to \tool.

{\bf \em Baselines.} We compare \tool with two specification-guided APR baselines, SpecRover~\cite{ruan2025specrover} and HoarePrompt~\cite{bouras2026hoareprompt}. SpecRover infers
natural-language intent specifications to guide patch generation and vet
candidate patches in a repository-level agent. Since SpecRover is 
designed for repo-level repair in AutoCodeRover, we adapt it to a file-level
variant, by~disabling its codebase search and navigation components, as {\tool} does not target repo-level bug fixing. HoarePrompt checks the correctness of a program by prompting an LLM to reason about whether it satisfies natural-language Hoare-style specifications. We adapt it to
program repair by keeping its specification-based reasoning
procedure and changing the final query from checking if the current program
is buggy to repairing the buggy program. All baselines are attached to multiple
repair backbones/settings, allowing us to evaluate them under one-turn,
iterative, and agentic file-level repair frameworks/settings.

{\bf \em Backbone Frameworks.} We evaluate each repair signal under three
backbone frameworks/settings. The first is the {\bf one-turn} setting, where each method
takes the buggy program~and is allowed to produce one fixed program
without iterative~feedback. The second setting is
ChatRepair~\cite{xia2024automated}, an {\bf iterative repair} framework that uses test
outcomes from previous attempts as conversational feedback to improve later
patches. The third setting is
AgentCoder~\cite{huang2024agentcodermultiagentbasedcodegeneration}, an {\bf agentic
repair} framework in which multiple roles cooperate to generate and refine
patches. Since ChatRepair uses a test-driven feedback as an APR signal, we use it
as a baseline and denote it by {\em Tester}.



{\bf \em Model Backbones.} We use GPT-4o-mini and Qwen3-235B-A22B-Instruct-2507  for cost efficiency. We apply nucleus sampling and {\em generate 5 candidate fixes per buggy problem}.

\begin{table*}[t]
    \centering
    \small
    \captionsetup{font=footnotesize}
    \caption{Program-repair performance across backbone frameworks, backbone models, and specification-guided repair methods. APR = Average Pass Rate. IT = input tokens, OT = output tokens, and TT = total tokens.}
    \label{tab:repair_performance}
    \newcommand{\bluecell}[1]{\cellcolor{blue!5}#1}
    \begin{tabular}{lllccccccc}
        \toprule
        \textbf{Backbone}
        & \textbf{Model}
        & \textbf{Method}
        & \multicolumn{4}{c}{\textbf{Functional Repair Metrics}}
        & \multicolumn{3}{c}{\textbf{Tokens (Millions)}} \\
        \cmidrule(lr){4-7}
        \cmidrule(lr){8-10}
        & & & \textbf{Pass@1} & \textbf{Pass@3} & \textbf{Pass@5} & \textbf{APR}
        & \textbf{IT} & \textbf{OT} & \textbf{TT} \\
        \midrule

        \multirow{6}{*}{One-turn}
        & \multirow{3}{*}{GPT-4o-mini}
        & SpecRover 
        & 42.33 & 48.05 & 49.44 & 67.52 & 4.01 & 0.91 & 4.92 \\
        & & HoarePrompt 
        & 40.22 & 52.78 & 57.78 & 64.52 & 1.04 & 0.52 & 1.56 \\
        & & \bluecell{\textbf{\tool~(Ours)}}
        & \bluecell{\textbf{47.34}} & \bluecell{\textbf{56.17}} & \bluecell{\textbf{60.56}} & \bluecell{\textbf{70.41}}
        & \bluecell{3.11} & \bluecell{1.30} & \bluecell{4.41} \\
        \cline{2-10}

        & \multirow{3}{*}{Qwen3-235B-A22B Instruct}
        & SpecRover 
        & 48.78 & 67.84 & 72.78 & 57.71 & 4.07 & 0.96 & 5.13 \\
        & & HoarePrompt 
        & 72.56 & 82.61 & \textbf{87.78} & 79.59 & 1.12 & 0.68 & 1.80 \\
        & & \bluecell{\textbf{\tool~(Ours)}}
        & \bluecell{\textbf{73.78}} & \bluecell{\textbf{83.23}} & \bluecell{85.56} & \bluecell{\textbf{83.05}}
        & \bluecell{3.26} & \bluecell{1.52} & \bluecell{4.78} \\
        \midrule

        \multirow{8}{*}{ChatRepair}
        & \multirow{4}{*}{GPT-4o-mini}
        & Tester
        & 50.33 & 50.84 & 51.11 & 82.99 & 12.89 & 5.51 & 17.40 \\
        & & SpecRover 
        & 51.78 & 52.50 & 52.78 & 82.21 & 21.76 & 6.43 & 28.19 \\
        & & HoarePrompt 
        & 43.01 & 43.45 & 43.89 & 79.01 & 8.57 & 6.12 & 14.69 \\
        & & \bluecell{\textbf{\tool~(Ours)}}
        & \bluecell{\textbf{55.00}} & \bluecell{\textbf{55.95}} & \bluecell{\textbf{56.67}} & \bluecell{\textbf{84.97}}
        & \bluecell{19.80} & \bluecell{5.81} & \bluecell{25.61} \\
        \cline{2-10}

        & \multirow{4}{*}{Qwen3-235B-A22B Instruct}
        & Tester
        & 86.89 & 87.84 & 88.84 & 96.01 & 10.57 & 13.42 & 24.09 \\
        & & SpecRover 
        & 81.67 & 83.78 & 84.45 & 93.12 & 22.84 & 7.41 & 30.25 \\
        & & HoarePrompt 
        & 88.34 & 88.67 & 88.89 & 96.08 & 9.25 & 6.94 & 16.19 \\
        & & \bluecell{\textbf{\tool~(Ours)}}
        & \bluecell{\textbf{91.56}} & \bluecell{\textbf{91.67}} & \bluecell{\textbf{91.67}} & \bluecell{\textbf{97.59}}
        & \bluecell{21.02} & \bluecell{5.61} & \bluecell{26.63} \\
        \midrule

        \multirow{8}{*}{AgentCoder}
        & \multirow{4}{*}{GPT-4o-mini}
        & Tester
        & 42.89 & 46.44 & 47.78 & 67.30 & 33.12 & 2.07 & 35.19 \\
        & & SpecRover 
        & 42.56 & 46.89 & 47.78 & 68.25 & 12.47 & 2.95 & 15.42 \\
        & & HoarePrompt 
        & 41.22 & 45.94 & 47.78 & 65.65 & 12.52 & 2.83 & 15.35 \\
        & & \bluecell{\textbf{\tool~(Ours)}}
        & \bluecell{\textbf{45.56}} & \bluecell{\textbf{49.01}} & \bluecell{\textbf{50.00}} & \bluecell{\textbf{67.98}}
        & \bluecell{17.16} & \bluecell{3.98} & \bluecell{21.14} \\
        \cline{2-10}

        & \multirow{4}{*}{Qwen3-235B-A22B Instruct}
        & Tester
        & 62.56 & 74.22 & 76.67 & 78.18 & 35.67 & 2.96 & 38.64 \\
        & & SpecRover 
        & 48.67 & 64.56 & 75.56 & 54.41 & 11.95 & 3.63 & 15.58 \\
        & & HoarePrompt 
        & 42.00 & 69.56 & 80.00 & 43.56 & 9.51 & 5.93 & 15.44 \\
        & & \bluecell{\textbf{\tool~(Ours)}}
        & \bluecell{\textbf{64.45}} & \bluecell{\textbf{77.00}} & \bluecell{\textbf{81.12}} & \bluecell{\textbf{78.77}}
        & \bluecell{13.10} & \bluecell{3.83} & \bluecell{16.93} \\
        \bottomrule
    \end{tabular}

    \vspace{1mm}
\end{table*}

{\bf \em Datasets.} Since \tool does not target repository-level APR, we use file-level repair benchmarks. However, existing file-level APR benchmarks may be exposed to modern LLMs, raising contamination concerns. We therefore collect recent problems from CodeForces (https://codeforces.com/), a competitive programming platform with continuously updated tasks and executable tests, to build a newer and less likely contaminated file-level benchmark. Each repair instance consists of a problem statement, a human-submitted buggy solution, and the corresponding hidden-quality test suite. Each problem is attached with a test suite ranging from 50 test cases to more than 200 test cases. We chose the buggy code with the cyclomatic complexity of 8 or more and has at least 15 LOCs. Finally, we obtained 180 buggy solutions across 90 problems.

\section{Program Repair Performance (RQ1)}
\label{sec:repair_perform}

\newcommand{\inc}[1]{\textsuperscript{\scriptsize $\uparrow$#1}}

\subsubsection{Settings} Each buggy program is subjected to {\em 5 independent repair attempts}. In the  one-turn setting, each attempt generates 5 candidate patches, from which the highest-performing one is selected as the final output. For iterative refinement setting, we set the default maximum number of refinement iterations to 20 for each attempt. With \tool integrated into them, the number of specification regeneration attempts is limited to 5 when trivial specifications are encountered.

\subsubsection{Metrics} We evaluate performance using \textit{Pass@k}. This metric measures whether at least one of the top-$k$ generated, fixed programs successfully {\em passes the entire test suite}. 


\subsubsection{Empirical Results} As seen in Table~\ref{tab:repair_performance}, in the {\bf \em one-turn} setting, \tool consistently improves over SpecRover and HoarePrompt, improving Pass@1 
5.0\% and 7.1\%, respectively with GPT-4o-mini and 25.0\% and 1.2\%, respectively with Qwen3-235B. Moreover, when integrated into {\bf \em iterative refinement} framework, \tool continues to deliver consistent improvements over Tester, SpecRover, and HoarePrompt 4.7\%, 3.2\%, and 12.0\%, respectively with GPT-4o-mini.
With the integration to the {\bf \em AgentCoder agentic APR framework}, {\tool} improves over Tester, SpecRover, and HoarePrompt 1.9\%, 10.0\%, and {22.5\%, respectively with Qwen3-235B.




\begin{figure*}[t]
 \centering
    \begin{minipage}[b]{0.48\textwidth}
    \centering
\begin{lstlisting}[language=Python, basicstyle=\scriptsize\ttfamily, numbersep=4pt, xleftmargin= 1pt]
# SpecTune
a,b,c,d=map(int,input().split())
o=max(a,b,c,d)
# assert o == max(a, b, c, d), ensuring the maximum is correctly identified
# assert o >= a and o >= b and o >= c and o >= d
if (a>=b and a>=c and a<b+c) or (a>=b and a>=d and a<b+d) or (a>=d and a>=c and a<c+d) or (b>=a and b>=c and b<a+c) or (b>=a and b>=d and b<a+d) or (b>=c and b>=d and b<c+d) or (c>=a and c>=d and c<a+d) or (c>=a and c>=b and c<a+b) or (c>=b and c>=d and c<b+d) or (d>=a and d>=b and d<a+b) or (d>=a and d>=c and d<a+c) or (d>=c and d>=b and d<c+b):
    # assert any( p + q > r for (p,q,r) in [ sorted(t) for t in [(a,b,c), (a,b,d), (a,c,d), (b,c,d)] ] )
    print('TRIANGLE')
elif 2*o<a+b+c+d:
    # assert not any( p + q > r for (p,q,r) in [ sorted(t) for t in [(a,b,c), (a,b,d), (a,c,d), (b,c,d)] ] )
    # assert any( p + q == r for (p,q,r) in [ sorted(t) for t in [(a,b,c), (a,b,d), (a,c,d), (b,c,d)] ] )
    print('SEGMENT')
else:
    # assert not any( p + q > r for (p,q,r) in [ sorted(t) for t in [(a,b,c), (a,b,d), (a,c,d), (b,c,d)] ] )
    # assert not any( p + q == r for (p,q,r) in [ sorted(t) for t in [(a,b,c), (a,b,d), (a,c,d), (b,c,d)] ] )
    print('IMPOSSIBLE')
\end{lstlisting}
\vspace{-15pt}
\captionsetup{font=footnotesize}`
\caption{Output from {\tool}: the condition at line 11 was selected by $\alpha$-$\beta$ algorithm: $\alpha$=1.0 (18 satisfies, 18 reaches, 67 passing tests), $\beta$=0.0 (0 satisfies, 7 reaches, 16 failing tests). It helps detect the (line 9) in that branch.}
\label{fig:spectune}
\end{minipage}
\hspace{5pt}
\begin{minipage}[b]{0.48\textwidth}
    \centering
\begin{lstlisting}[language=Python, basicstyle=\scriptsize\ttfamily, numbersep=4pt]
# HOAREPROMPT
# State of the program right berfore the function call: stdin contains one input: four space-separated positive integers, each not exceeding 100, ...
a, b, c, d = map(int, input().split())
o = max(a,b,c,d)
if (a >=b and a >= c and a <b+c or a>=b and a>=d and a<b+d or a>=d and a>=c and a<c+d or b>=a and b>=c and b<a+c or b>=a and b>=d and b<a+d or b>=c and b>=d and b<c+d or c>=a and c>=d and c<a+d or c>=a and c>=b and c<a+b or c>=b and c>=d and c<b+d or d>=a and d>=b and d<a+b or d>=a and d>=c and d<a+c or d>=c and d >= b and d<c+b):
    print('TRIANGLE')
    #This is printed: TRIANGLE
elif (2*o <a+b+c+d) :
       print('SEGMENT')
       #This is printed: SEGMENT
else :
       print('IMPOSSIBLE')
       #This is printed: IMPOSSIBLE
       #State: a, b, c, d are positive integers not exceeding 100, representing the lengths of four sticks, and o is the maximum value among a, b, c, and d. It is guaranteed that no triplet of sticks can form a triangle, meaning for every triplet where one stick is the longest,...
# Overall this is what the function does: The function reads four space-separated positive integers ...
\end{lstlisting}
\vspace{-15pt}\captionsetup{font=footnotesize}`
\caption{Output from HoarePrompt~\cite{bouras2026hoareprompt}: The bug at line 9 with that branch was not detected by HoarePrompt as its generated condition at line 11: "This is printed: SEGMENT", which is not executable by test cases.}
\label{fig:hoareprompt}
\end{minipage}
\end{figure*}

\subsubsection{Qualitative analysis} We found that our ($\alpha$,$\beta$) algorithm
selects high-quality condition signals, helping code repair, better than
natural-language specifications. Let us illustrate in Figs~\ref{fig:spectune}-~\ref{fig:hoareprompt}. The executable condition at line 11 from {\tool}
was selected by our algorithm as $\alpha$=1.0 (correctness) and $\beta$=0 (completeness), leading to {\tool} detecting the bug in that branch. In contrast, HoarePrompt simply produced {\em "This is printed: SEGMENT"}, which was not useful here. By using feedback about trivial or unreliable specifications from previous refinement turns, with the results of passing and failing tests, the model can generate better specifications in later turns. Moreover, the condition from {\tool} at line 7 has 30 passing tests, all 30 satisfied, while none of the failing test cases went through. That could signal the model that the error might be elsewhere.

\subsubsection{Token Usage Efficiency} 
We measure token usage by the number of input tokens, output tokens, and total tokens consumed when integrating specification-guided repair into existing repair frameworks. 
Table~\ref{tab:repair_performance} reports input tokens (IT), output tokens (OT), and total tokens (TT) in millions. 
Since refinement-based and agentic frameworks involve iterative feedback and multiple repair attempts, their token usage is naturally higher than the one-turn setting. In the one-turn setting, \tool{} uses 4.41M total tokens with GPT-4o-mini and 4.78M total tokens with Qwen3-235B. 
These numbers are comparable to SpecRover, which uses 4.92M and 5.13M total tokens, while being higher than HoarePrompt, which uses 1.56M and 1.80M total tokens. For iterative and agentic frameworks, \tool{} also maintains competitive token usage compared with the baselines, using 25.61M/26.63M total tokens in ChatRepair and 21.14M/16.93M total tokens in AgentCoder under GPT-4o-mini/Qwen3-235B, respectively. 

Overall, \tool{} introduces token usage comparable to existing specification-guided repair methods, while achieving better functional quality across backbone frameworks.

\section{Fault Localization (FL) Capability (RQ2)}
\label{sec:rq2}

One of the key design goals of {\tool} (\textbf{Key Idea 1}) is to provide
specification-guided signals that improve the reasoning capability of
LLMs, enabling them {\em to more effectively narrow the search space and
identify error-prone code regions}, leading to better locating faulty lines. 
Thus, we compare the localization performance of SpecRover and HoarePrompt using each of the two backbone LLMs with that of {\tool} with the same model under one-turn setting. That also helps to explain the high APR performance of {\tool} as seen in RQ1.

\subsubsection{Settings} Since the dataset does not include
explicit ground-truth buggy locations, we evaluate fault localization
using an LLM-as-a-judge paradigm. For each bug instance, each approach under evaluation predicts {\color{black}{suspicious line(s)}}. These predictions are then provided to a judge LLM, which performs a
pairwise comparison and returns a binary preference indicating which
prediction better captures the {\color{black}{faulty line(s)}}. For the judge model, to improve evaluation robustness and reduce model-specific bias, we use \textsc{GPT-5.2} and \textsc{GPT-5.4}, leveraging their strong reasoning capabilities across diverse code-related tasks. To further ensure the reliability of the evaluation, we retain only the subset of instances where all judge models are able to produce a correct repair, indicating their capability to accurately reason about the underlying bug. In addition, we also manually inspect a sampled subset of the localization results for human evaluation. To assess the consistency between human judgments and LLM-based judgments, we report their inter-rater agreement using  Cohen's $\kappa$ (for two raters): $\kappa = \frac{p_o - p_e}{1 - p_e}$, where $p_o$ denotes the observed agreement between the two raters, and $p_e$ denotes the expected agreement by chance.

\subsubsection{LLM-judge Metrics}
Let $\mathcal{D}$ denote the set of bug instances where both a baseline (SpecRover or HoarePrompt) and {\tool} produce valid localization predictions. For each instance $i \in \mathcal{D}$, let $s_i$ and $n_i$ denote the sets of predicted faulty lines from {\tool} and the base LLM. 
We employ an LLM-based judge function $\mathrm{Judge}(s_i, n_i)$, which takes two sets of predicted lines as input and returns a binary outcome: $0$ if $s_i$ is preferred over $n_i$, and $1$ otherwise. We define the following metrics:
$
\text{\bf WinRate}(\text{\tool}) = \frac{|\{i \in \mathcal{D} \mid \mathrm{Judge}(s_i, n_i) = 0\}|}{|\mathcal{D}|}
$, 
$
\text{\bf WinRate}(\text{Base})$=$\frac{|\{i \in \mathcal{D} \mid \mathrm{Judge}(s_i, n_i) = 1\}|}{|\mathcal{D}|}
$,
$
\text{\bf DrawRate}$ = $\frac{|\{i \in \mathcal{D} \mid s_i = n_i\}|}{|\mathcal{D}|}
$, where $\text{WinRate}(\text{\tool})$ measures the proportion of instances where the judge prefers the result from {\tool}, while $\text{WinRate}\-(\text{Base})$ captures the opposite cases. $\text{DrawRate}$ corresponds to instances where both methods predict identical sets of buggy lines.

\subsubsection{Empirical Results} 
We conduct the experiment on 119 tasks on our dataset for GPT-4o-mini and Qwen3-235B, where violated specifications are successfully generated by
the baselines under the condition $\alpha \geq 0.9$ and $\beta < 1$. 
Table~\ref{tab:FL} presents the overall impact of the generated postconditions on fault localization. 
Overall, {\tool} consistently outperforms the baselines across all settings, demonstrating that the generated postconditions, guided by the estimated $\alpha$ and $\beta$ signals, are effective in improving localization quality. 

When using GPT-5.2 as the judge, {\tool} first shows substantial improvements over the naive baseline, winning in 
107 cases (89.91\%) under GPT-4o-mini and 
114 cases (95.79\%) under Qwen3-235B. 
Compared with stronger natural-language specification baselines, {\tool} achieves 58 cases (48.73\%) and 59 cases (49.58\%) where its predictions are preferred over SpecRover and HoarePrompt, respectively, under GPT-4o-mini. A similar trend is observed for Qwen3-235B, where {\tool} is preferred in 47 cases (39.50\%) and 48 cases (40.34\%) over SpecRover and HoarePrompt, respectively. 

When using GPT-5.4 as the judge, {\tool} remains consistently preferred over the baselines. Against the naive baseline, {\tool} wins in 
102 cases (85.72\%) under GPT-4o-mini and 
113 cases (94.95\%) under Qwen3-235B. 
Under GPT-4o-mini, {\tool} further wins over SpecRover and HoarePrompt in 
57 cases (47.90\%) and 52 cases (43.71\%), respectively. Under Qwen3-235B, {\tool} is preferred in 
39 cases (32.78\%) and 38 cases (31.93\%) 
over SpecRover and HoarePrompt, respectively.
Under GPT-5.4 as the judge, the results remain consistent. 
That is, {\tool} yields a higher proportion of preferred outcomes than the baselines,~regardless of the underlying model or judge.


\subsubsection{Qualitative Analysis} 
{\tool} provides a fine-grained view of where the buggy program begins to diverge from the intended behavior. The signal comes from two sources: {\bf \em whether an execution reaches a checkpoint}, and {\bf \em whether the corresponding postcondition is satisfied} when it is reached. If passing and failing executions share earlier checkpoints but the failing execution first violates a later postcondition, the suspicious region can be narrowed to the code segment leading to that checkpoint. If a failing execution fails to reach a checkpoint that is normally reached by passing executions, the signal suggests that the bug may occur elsewhere in the control flow. In Fig.~\ref{fig:spectune}, the assertion on line 7 has 30 passing tests going through; all
30 were satisfied, yet no failing test cases reached it. This helps the model look elsewhere.
In our experiments, these dynamic checkpoint-level signals help the model reason more precisely about program states.



\begin{table}[t]
\centering
\captionsetup{font=footnotesize}
\caption{{Pairwise fault-localization comparison between \tool{} and natural-language specification baselines.}}
\label{tab:FL}
\small
\setlength{\tabcolsep}{3.5pt}
\vspace{-2pt}
\renewcommand{\arraystretch}{0.98}
\begin{tabular}{@{}llccc@{}}
\toprule
\textbf{Judge} & \textbf{Comparison} 
& \textbf{\tool{}} & \textbf{Baseline} & \textbf{Draws} \\
\midrule

\rowcolor{gray!18}
\multicolumn{5}{c}{\textbf{GPT-4o-mini}} \\
GPT-5.2 & \tool{} vs Naive       & \textbf{89.91} & 4.21  & 5.88  \\
GPT-5.2 & \tool{} vs HoarePrompt & \textbf{49.58} & 29.42 & 21.00 \\
GPT-5.2 & \tool{} vs SpecRover   & \textbf{48.73} & 31.09 & 20.18 \\
GPT-5.4 & \tool{} vs Naive       & \textbf{85.72} & 8.40  & 5.88  \\
GPT-5.4 & \tool{} vs HoarePrompt & \textbf{43.71} & 35.29 & 21.00 \\
GPT-5.4 & \tool{} vs SpecRover   & \textbf{47.90} & 31.92 & 20.18 \\

\midrule
\rowcolor{gray!18}
\multicolumn{5}{c}{\textbf{Qwen3-235B-A22B Instruct}} \\
GPT-5.2 & \tool{} vs Naive       & \textbf{95.79} & 3.36  & 0.85  \\
GPT-5.2 & \tool{} vs HoarePrompt & \textbf{40.34} & 17.65 & 42.01 \\
GPT-5.2 & \tool{} vs SpecRover   & \textbf{39.50} & 22.69 & 37.81 \\
GPT-5.4 & \tool{} vs Naive       & \textbf{94.95} & 4.20  & 0.85  \\
GPT-5.4 & \tool{} vs HoarePrompt & \textbf{31.93} & 26.06 & 42.01 \\
GPT-5.4 & \tool{} vs SpecRover   & \textbf{32.78} & 29.41 & 37.81 \\

\bottomrule
\end{tabular}
\end{table}


\subsubsection{Human Evaluation} \normalfont
To further assess the reliability of LLM-as-a-judge, we conduct a human study on a randomly sampled subset of 30 bug instances. For each instance, human are asked to compare the localization outputs produced by the baselines and {\tool}, and determine which prediction better captures the faulty region. To quantify the agreement between human judgments and LLM-based decisions, we compute Cohen's $\kappa$ between human annotations and each LLM judge.
As seen in Table~\ref{tab:human_agreement}, Cohen's $\kappa$ indicates substantial agreement between humans and GPT-5.2 ($\kappa$=0.629) and GPT-5.4 ($\kappa$= 0.731). This further confirms that the judgments produced by the LLM are largely consistent with human assessments. 


\begin{table}[t]
\centering
\captionsetup{font=footnotesize}`
\caption{Agreement between human annotators and LLM judges measured by Cohen's $\kappa$.}
\label{tab:human_agreement}
\begin{tabular}{lcc}
\toprule
\textbf{Metric} & \textbf{Human vs GPT-5.2} & \textbf{Human vs GPT-5.4} \\
\midrule
Cohen's $\kappa$ & 0.629*** & 0.731*** \\
\bottomrule
\multicolumn{3}{l}{\footnotesize *** $p < 0.001$}
\end{tabular}
\end{table}

\subsubsection{\bf \em Conclusion} These suggest that using specification-guided signals in \tool effectively helps narrow down the search space for fault localization. This also reinforces the findings of RQ1 (Section \ref{sec:repair_perform}) and {\bf Key Idea 1} (Section~\ref{sec:key_ideas}).


\section{Effectiveness of ($\alpha$-$\beta$) Algorithm (RQ3)}
\label{sec:rq3}

$\alpha$-$\beta$ algorithm aims to retain high-quality LLM-generated conditions. This RQ evaluates it from two perspectives. First, we assess whether the estimated consistency signal $\alpha$, computed from executions of the buggy program, reliably approximates the consistency of the same postconditions on the fixed program. Second, we analyze how effectively the full $\alpha$-$\beta$ filtering algorithm removes inconsistent or uninformative postconditions, retaining the high-quality specifications for APR.



\subsection{Reliability of the $\alpha$ Consistency Estimate}
\label{ab-quality}

\subsubsection{Settings} {\tool} generates checkpoint-level postconditions from the buggy code and the problem description. These postconditions are intended to represent hypotheses about the expected program state, analogous to how a developer reasons about intermediate behavior during debugging. To evaluate whether our estimated consistency signal $\alpha$ is reliable, we need access to the corresponding fixed program and a mapping from checkpoints in the buggy program to locations in the fixed program. Because our main dataset does not provide human-written fixes, we use HumanEvalFix~\cite{muennighoff2024octopackinstructiontuningcode}, which contains 168 coding problems and provides buggy/fixed program pairs. However, HumanEvalFix contains only 7.25 test cases per problem on average, which is insufficient for reliably estimating postcondition consistency. Therefore, we use GPT-4.1-mini to generate additional inputs aimed at increasing branch coverage. We execute these inputs on the canonical solution to obtain additional test cases and merge them with the original tests. Overall, each task has 102.38 test cases on average.


\subsubsection{Procedure} Given a buggy program $C$, we run {\tool} to generate postconditions at multiple checkpoints and compute their estimated consistency signal $\alpha$ using only executions of the buggy program, as done in the normal {\tool}~workflow. We then map each checkpoint to its respective location in the fixed program $C*$ and execute $C*$ on the tests. Based~on whether the same postconditions hold at the mapped checkpoints in $C*$, we compute a reference consistency signal $\alpha^{*}$.

{\em By comparing $\alpha$ with $\alpha^*$, we evaluate how well the consistency signal estimated from the buggy program approximates the reference consistency measured on the fixed program}. This experiment focuses only on the reliability of the $\alpha$ estimate; the effectiveness of the full $\alpha$-$\beta$ algorithm in retaining high-quality postconditions is evaluated in Section~\ref{sec:spec-quality}.



\begin{figure}
    \centering
    \includegraphics[width=0.6\linewidth]{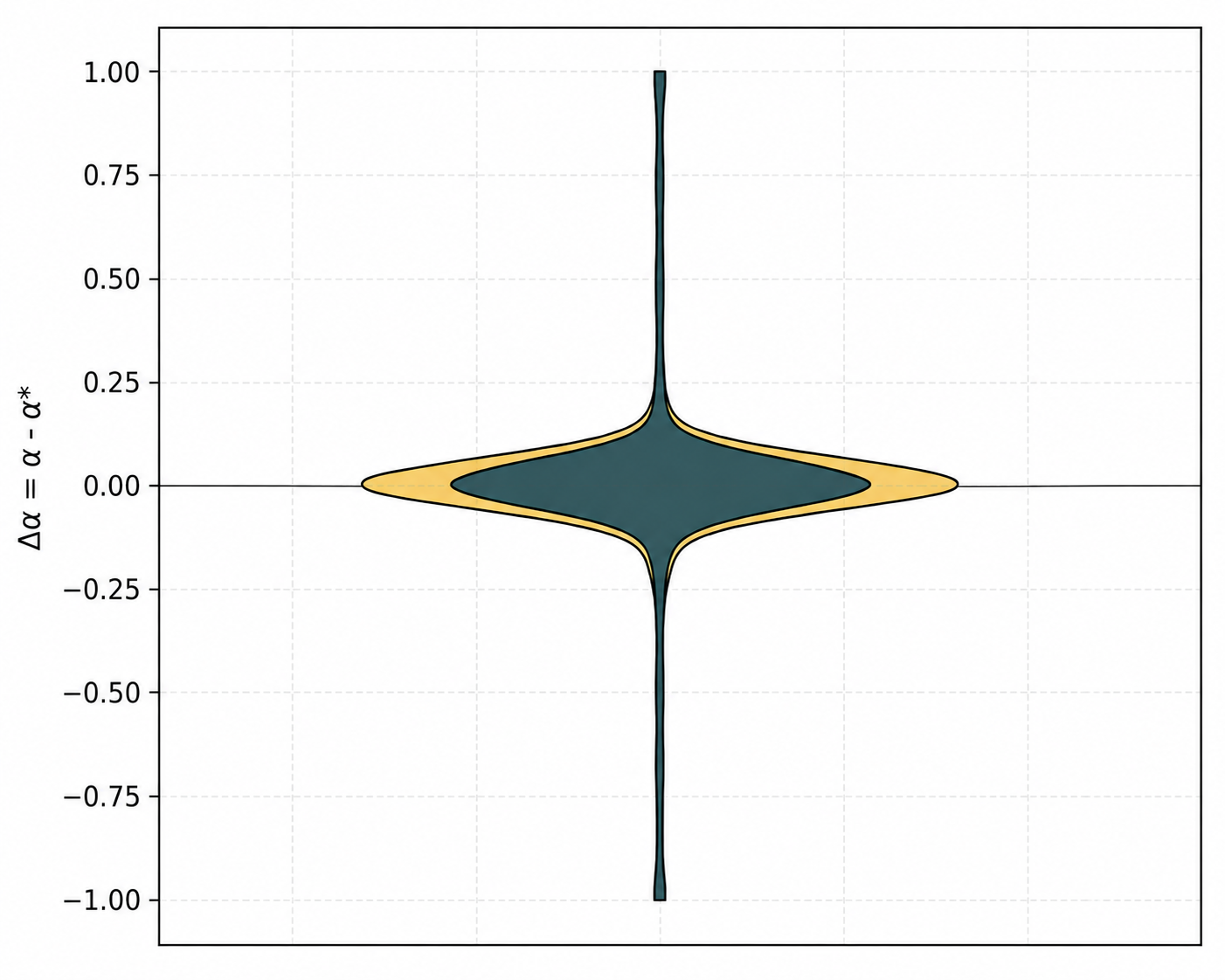}
    \vspace{-3pt}
    \captionsetup{font=footnotesize}`
    \caption{{\color{custom-blue}Distribution of $\Delta\alpha = \alpha - \alpha^{*}$ across all postconditions from GPT-4o-mini (yellow) and Qwen3-235B (blue)}. }
    \label{fig:placeholder}
\end{figure}

\subsubsection{Empirical Results} Fig.~\ref{fig:rq3} presents the distribution of 
$\Delta\alpha = \alpha - \alpha^{*}$ across all generated postconditions for both backbone  models. In our evaluation, we consider a postcondition to be \emph{highly consistent} with the reference signal if $|\Delta\alpha| < 0.1$, meaning the estimated consistency signal deviates from the true reference by less than 0.1. The plot reveals that the distribution is sharply concentrated around zero for both models: {\em over 78.6\% of postconditions generated by GPT-4o-mini and over 81.2\% of those generated by Qwen3-235B are classified as highly consistent}. Furthermore, {\em both models yield near-zero mean and median errors}: GPT-4o-mini achieves a mean $\Delta\alpha$ of 0.004 with a median of 0.000, while Qwen3-235B obtains a mean of -0.02 and a median of 0.000. 

\subsubsection{\bf \em Analysis} 
These results show that $\alpha$ is a reliable estimator of postcondition consistency. Despite relying only on executions of the buggy program, $\alpha$ closely approximates the reference consistency measured on the fixed code. This confirms \textbf{Key Idea 3}: $\alpha$ provides a dependable validation signal for assessing and filtering LLM-generated postconditions.

\subsection{Distribution of Postcondition Quality}
\label{sec:spec-quality}

\begin{figure}
    \centering
    \includegraphics[width=0.7\linewidth]{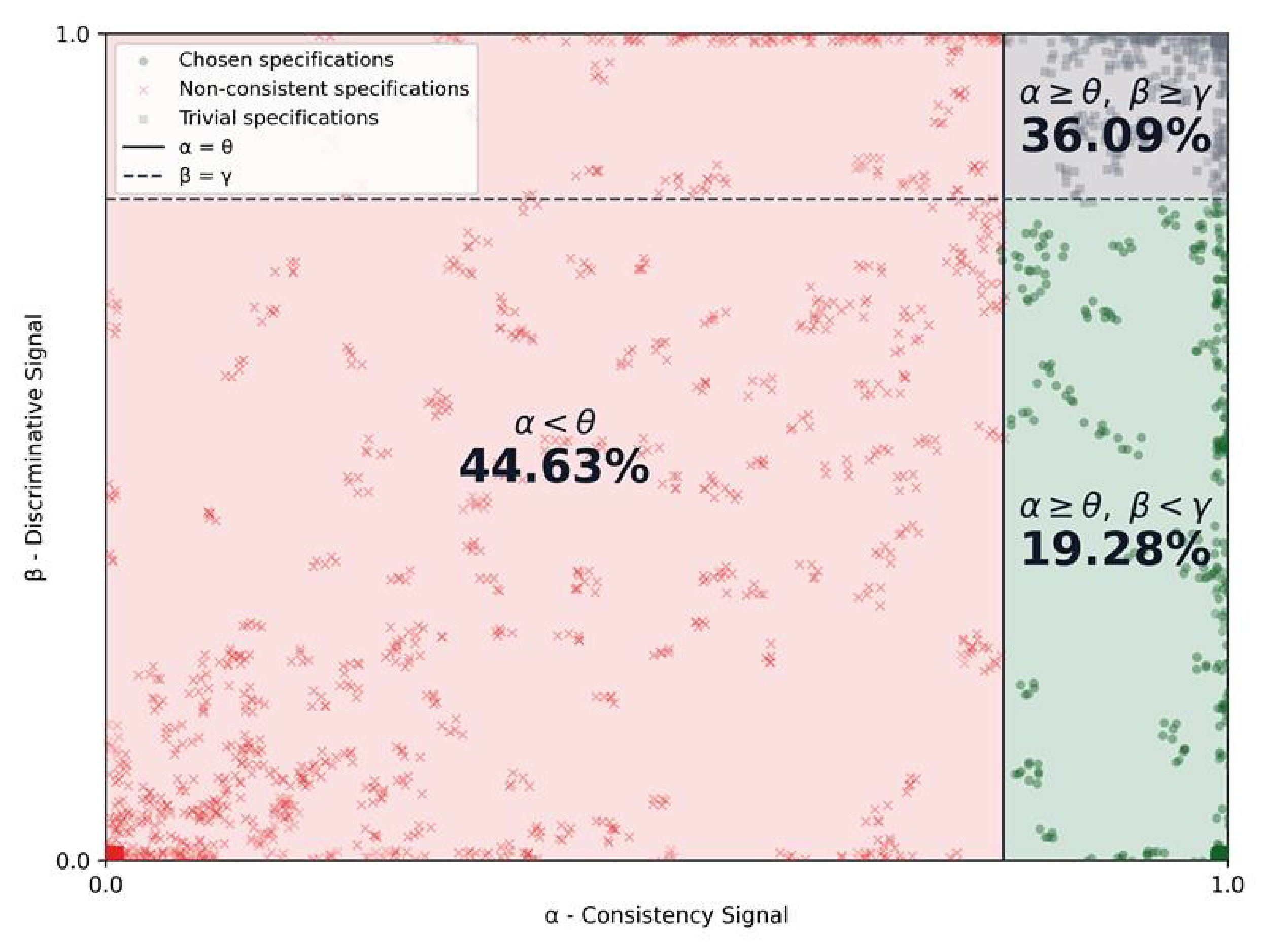}
    \captionsetup{font=footnotesize}`
    \caption{{\color{custom-blue}Distribution of generated condition quality by GPT-4o-mini, estimated using the $\alpha$ consistency signal (x-axis) and the $\beta$ discriminative signal (y-axis)}.}
    \label{fig:rq3}
\end{figure}

This section analyzes the quality distribution of the postconditions
generated by the LLM at multiple checkpoints. Having established in
Section~\ref{ab-quality} that the estimated signal $\alpha$ reliably approximates
the consistency quality of generated postconditions, we now apply
{\tool} to the buggy programs in our dataset and compute the
$\alpha$ and $\beta$ values of the generated checkpoint-level
postconditions. We then categorize each postcondition into one of
three regions: the \textbf{\textcolor{NonConsistent}{non-consistent region}} if $\alpha(s) < \theta$; the \textbf{\textcolor{Trivial}{trivial region}} if $\alpha(s) \ge \theta$ and $\beta(s) \ge \gamma$; and the \textbf{\textcolor{Acceptable}{acceptable region}} if $\alpha(s) \ge \theta$ and $\beta(s) < \gamma$.  Each region reflects a different level of specification quality with respect to consistency and the ability to detect incorrect behavior. This experiment is designed to characterize {\em how often LLM-generated
postconditions are usable for repair, and to quantify the extent to
which inconsistent or trivial specifications must be filtered out
before they are used as debugging signals}.



Fig.~\ref{fig:rq3} shows the distribution of generated specification quality by GPT-4o-mini, using $\theta$ = $\gamma$ = 0.80. The \textbf{\textcolor{Trivial}{trivial region}} has the highest density at 36.09\% and the \textbf{\textcolor{NonConsistent}{non-consistent region}} accounts for 44.63\%, indicating that the model often generates specifications with little practical value. The plot also shows that specifications are highly concentrated around $(\alpha, \beta) = (0, 0)$, corresponding to inconsistent cases, and $(1, 1)$, corresponding to trivial ones. Only 19.28\% of specifications fall into the \textbf{\textcolor{Acceptable}{acceptable region}}. We present the density of each region across multiple threshold values in Table~\ref{tab:spec_distribution}. 



\begin{table}[t]
 \centering
 \captionsetup{font=footnotesize}`
 \caption{{\color{custom-blue}Distribution of generated specification quality by GPT-4o-mini across different threshold values $\theta$ and $\gamma$. Percentages indicate the proportion of specifications in each region.}}
 \small
 \label{tab:spec_distribution}
 \tabcolsep 3pt
 \vspace{6pt}
 \begin{tabular}{ccccc}
 \toprule
 $\theta$ & $\gamma$ & \textbf{\textcolor{NonConsistent}{Non-Consist. (\%)}} & \textbf{\textcolor{Trivial}{Trivial (\%)}} & \textbf{\textcolor{Acceptable}{Acceptable (\%)}} \\
 \midrule
 0.90 & 0.90 & 45.99 & 34.66 & 19.36 \\
 0.90 & 1.00 & 45.99 & 33.55 & 20.47 \\
 1.00 & 0.90 & 48.61 & 33.23 & 18.16 \\
 1.00 & 1.00 & 48.61 & 36.84 & 12.76 \\
 \bottomrule
\end{tabular}
\end{table}

\subsubsection*{\bf \em Analysis} The results have two implications. \underline{First}, we do not need many correct specifications for APR. While the overall yield of acceptable specifications is low, this does not hinder the effectiveness of {\tool} as one high-quality that can detect the fault is needed. From a human debugging perspective, a programmer does not attempt to generate as many specifications as possible; instead, they brainstorm only a small subset of specifications that capture the most suspicious or informative program states at relevant checkpoints. This targeted reasoning is sufficient to localize and fix a bug. 
\underline{Second}, the results also confirm our {\bf Key Ideas 2 and 3} on the need for a filtering mechanism to prevent low-quality specifications from being used in code repair. 
Our ($\alpha$-$\beta$) algorithm ensures that only consistent and discriminative specifications are retained for repair. As shown in the ablation study (Table~\ref{tab:rq4_ablation}), removing the ($\alpha$-$\beta$) algorithm significantly degrades performance, as a large number of inconsistent or trivial specifications are introduced into the repair process, misleading the fault localization and reducing the overall bug-fixing effectiveness.






\section{Ablation Study (RQ4)}
\label{sec:rq4}
This section examines the contribution of each signal $\alpha$ and $\beta$. Specifically, we analyze how removing the consistency signal $\alpha$ and the discriminative signal $\beta$ affects its APR effectiveness.

\begin{table}[t]
\captionsetup{font=footnotesize}
\caption{Effect of $\alpha$ and $\beta$ signals on \tool performance.}
\label{tab:rq4_ablation}
\centering
\small
\vspace{-3pt}
\setlength{\tabcolsep}{4pt}
\renewcommand{\arraystretch}{0.98}
\begin{tabular}{@{}lcccc@{}}
\toprule
\textbf{Setting} & \textbf{Pass@1} & \textbf{Pass@3} & \textbf{Pass@5} & \textbf{APR} \\
\midrule
\textbf{Full SpecTune}            & \textbf{47.34} & \textbf{56.17} & \textbf{60.56} & \textbf{70.41} \\
w/o Consistency Signal $\alpha$   & 44.89 & 53.17 & 55.56 & 69.72 \\
w/o Discriminative Signal $\beta$ & 45.67 & 53.78 & 57.23 & 69.57 \\
w/o $\alpha$ and $\beta$ signals  & 44.11 & 52.34 & 55.56 & 69.49 \\
\bottomrule
\end{tabular}
\end{table}

\begin{table}[t]
\centering
\footnotesize
\captionsetup{font=footnotesize}
\caption{Sensitivity analysis on quality thresholds.}
\label{tab:rq4_sensitivity}
\small
\begin{tabular}{cc|cccc}
\toprule
\textbf{$\alpha \geq \theta$} & \textbf{$\beta < \gamma$} 
& \textbf{Pass@1} & \textbf{Pass@3} & \textbf{Pass@5} & \textbf{APR} \\
\midrule

\multirow{3}{*}{0.9}
& 0.9  & 46.44 & 56.12 & 60.00 & 71.49 \\
& 0.95 & 45.78 & 56.12 & 57.78 & 69.69 \\
& 1.0  & 47.34 & 56.17 & 60.56 & 70.41 \\
\midrule

\multirow{3}{*}{0.95}
& 0.9  & 44.67 & 54.16 & 57.89 & 68.35 \\
& 0.95 & 45.44 & 51.56 & 55.00 & 69.28 \\
& 1.0  & 45.78 & 54.45 & 58.66 & 70.08 \\

\midrule

\multirow{3}{*}{1.0}
& 0.9  & 45.00 & 53.89 & 58.89 & 68.61 \\
& 0.95 & 45.67 & 53.55 & 56.12 & 70.82 \\
& 1.0  & 45.22 & 53.89 & 57.21 & 69.82 \\

\bottomrule
\end{tabular}
\end{table}


\subsubsection{Setting} \normalfont We conduct an ablation study of  under the one-turn setting when {\tool} integrated with the base LLMs from RQ1 (Section ~\ref{sec:repair_perform}) by systematically removing each component: (1) removing the consistency signal $\alpha$ (i.e., keeping all specifications that satisfy $\beta$<1 regardless of their consistency with passing executions), (2) removing the discriminative signal $\beta$ (i.e., retaining all specifications that satisfy $\alpha \ge 0.9$ regardless of their ability to detect faulty behaviors), and (3) removing both signals (i.e., using all generated specifications without any filtering). As both underlying models exhibit consistent trends across all three RQs above, for brevity, we evaluate each ablated variant on GPT-4o-mini only.

\subsubsection{Empirical Results} \normalfont 
Table~\ref{tab:rq4_ablation} shows that removing either signal degrades performance across all metrics. Removing $\alpha$ alone drops Pass@1 by 2.45\%, this suggests that without~consistency filtering, inconsistent postconditions are passed to the repair process, misleading the Repair Model toward incorrect reasoning about the desired program behavior. Otherwise, removing $\beta$ alone results in a less drop (1.67\%) in Pass@1, showing that keeping high-quality specification is crucial. This suggests that introducing trivial postconditions into the repair process reduces the LLM's ability to reason about the root cause of the bug. Finally, we removed both signals, meaning that all postconditions generated by the LLM are passed directly into the repair process. The results show that this leads to the largest degradation, with a 3.23\% drop in Pass@1.



\section{Sensitivity Analysis (RQ5)}

This section aims to investigate the sensitivity of \tool to its key hyperparameters: the consistency signal $\alpha$ and the discriminative signal $\beta$, with respect to its performance. 


\subsubsection{Setting} \normalfont To ensure both consistency with correct executions and correctness in identifying faults, we consider threshold ranges for the consistency and discriminative signals as $\theta, \gamma \in \{0.9, 0.95, 1.0\}$, where specifications are retained if $\alpha \geq \theta$ and $\beta < \gamma$. Similar to RQ4 (Section ~\ref{sec:rq4}), we perform this experiment using the \tool configuration integrated with backbone LLMs, as established in RQ1 (Section ~\ref{sec:repair_perform}) and report only the results for Qwen3-235B only.

\subsubsection{Empirical Results} \normalfont Table~\ref{tab:rq4_sensitivity} presents the sensitivity of \tool to the thresholds $\theta$ and $\gamma$. Overall, \tool demonstrates robust and consistent performance across all evaluated hyperparameter combinations, with all variants consistently outperforming configurations without specification filtering. In our study, the configuration $(\theta, \gamma) = (0.9, 1.0)$ yields the best overall results. From these observations, we notice that for a fixed $\theta$, increasing $\gamma$ toward $1.0$ - which means we only filter out postconditions that are \emph{completely trivial} with $\beta$=1, consistently yields equal or better performance. This indicates that specifications with {\em discriminative signal $\beta \in [0.9, 1.0)$ still carry useful fault-localizing and APR information}. 
\section{Threats to Validity}
\label{sec:limit_threats}

\emph{Internal Validity.}  {\tool} depends on LLM-generated postconditions, whose quality may 
suffer from hallucinations. We mitigate this threat through our ($\alpha$-$\beta$) mechanism. Another threat is the nondeterminism of LLM-based generation; to reduce its impact, we use the same
backbones and controlled settings when comparing baseline and
augmented variants.

\emph{External Validity.}
Results may not generalize to other languages, large systems, or domain bugs.

\emph{Construct Validity.}
Test-suite pass rates may not fully capture correctness due to its incompleteness. 

\section{Related Work}
\label{sec:related_work}

In addition to classic APR~\cite{weimer2009genprog,pacheco2019prapr,liu2019tbar,liu2019avatar,nguyen2013semfix,mechtaev2015sketchfix,chen2021jaid,koyuncu2018fixminer} and deep learning APR methods~\cite{lutellier2020coconut,ye2021cure,zhu2021recoder,ye2023selfapr,tang2023tare,li2023knod} (Section~\ref{sec:intro}), we focus on LLM-based APR approaches.




{\bf \emph{Single-pass, non-iterative LLM-based APR.}}
Recent advances in LLMs have transformed APR
by enabling powerful zero-shot and few-shot patch generation.
Prior studies~\cite{fan2023apr_llm_outputs,xia2023plm_apr} show that large pretrained models can
generate plausible fixes when prompted with buggy code and context.
\textsc{AlphaRepair}~\cite{xia2022alpharepair} leverages
large pretrained code models to generate patches with infilling-style prediction, 
while \textsc{Repilot}~\cite{kang2023repilot}
integrates LLM-based code completion with patch validation.


{\bf \emph{LLM-based APR with iterative refinement.}} \textsc{ChatRepair}~\cite{xia2024automated} leverages conversational
feedback to iteratively improve patches based on test results, while
REx~\cite{tang2024code} formulates repair as a budgeted search problem
over candidate trajectories.
{\em Retrieval-augmented approaches} such as
RAP-Gen~\cite{wang2023rapgen} incorporate similar bug-fix exemplars into
the prompt.

Our experimental results show that {\bf {\tool} is able to improve the APR performance}
of the underlying LLMs in a single-pass and iterative repair settings.

{\bf \emph{Agentic APR frameworks.}}
Multi-agent frameworks, e.g., \textsc{AutoCodeRover}~\cite{zhang2024autocoderover}
and \textsc{RepairAgent}~\cite{bouzenia2025repairagent} treat APR as an
autonomous workflow, where LLMs iteratively localize faults, analyze
context, and invoke tools such as compilation and testing.
We {\bf did not compare} or add {\tool} to {\bf repository-level, multi-agent} APR systems, since they introduce extra confounding factors, including context retrieval, tool use,
planning, test generation, build setup, and patch ranking,
making it difficult to attribute improvements to {\tool}.


{\bf \emph{Specification generation and test synthesis.}}
Learning techniques have been explored for generating
program specifications and test artifacts, including postcondition
inference~\cite{nl2postcond, le-etal-2026-specmind}, test oracle synthesis~\cite{dinella2022toga,mastropaolo2023usingtransfer,10.1145/3524481.3527220},
test coverage improvement~\cite{lemieux2023codamosa,coverup2025,le2026testweaver}, and
unit test generation~\cite{lahiri2023interactivecodegenerationtestdriven,tufano2021unittestcasegeneration}.
EvoSpex~\cite{10.1109/ICSE43902.2021.00112} infers input--output relations,
while others generate property-based tests~\cite{vikram2024largelanguagemodelswrite}
or candidate properties~\cite{DBLP:journals/corr/abs-2210-00848}.

Machine learning has also been applied to infer specifications,
such as program invariants~\cite{10.1145/2837614.2837664,Laich2020Guiding,10.1145/3385412.3385986,pmlr-v202-pei23a}. SpecRover~\cite{ruan2025specrover} is an intent/specification-inference approach for improving the agentic APR framework AutoCodeRover. It iteratively infers natural-language specifications to guide patching, vet patches, and provide confidence/evidence. We have shown {\tool} performs better than SpecRover and HoarePrompt. NL2Postcond~\cite{nl2postcond} leverages LLMs to infer postconditions from textual descriptions of code. We do not compare with NL2Postcond because it targets a different setting. It generates method-level postconditions either from natural-language intent alone or from a correct/reference program. In contrast, \tool takes a buggy program and its natural-language description as input and generates checkpoint-level postconditions at intermediate locations. Thus, NL2Postcond does not directly match our input or repair-guidance setting.


\section{Conclusion}
\label{sec:conclusion}

This paper introduces \tool, a specification-guided framework for improving code LLMs in APR. By leveraging intermediate specifications and execution signals, \tool improves single-pass repair and can enhance existing iterative APR frameworks. Our results show consistent repair gains, improved fault localization, and more meaningful behavioral guidance, suggesting that specification-guided reasoning is a promising direction for reliable LLM-based APR.

{\bf Data Availability.} All data and code are available at~\cite{spectune}.



\balance

\bibliographystyle{IEEEtran}

\bibliography{references,apr-references,refs-specmining}

\end{document}